\documentclass[aps,twocolumn,superscriptaddress]{revtex4}
\usepackage{graphicx,graphics}
\usepackage{amsmath,epsf,subfigure}
\usepackage{color}
\usepackage{amsfonts,amsmath,graphics}
\usepackage{bm,subfigure}
\usepackage{graphicx,epsf}
\usepackage{amssymb}
\usepackage{afterpage}
\usepackage{float}
\usepackage{chemarr}
\usepackage{upgreek}
\usepackage{tikz-cd}
\usepackage{setspace}

\renewcommand{\r}{\rho}

\newcommand{\beq}{\begin{equation}}
\newcommand{\eeq}{\end{equation}}
\newcommand{\bea}{\begin{eqnarray}}
\newcommand{\eea}{\end{eqnarray}}
\newcommand{\harp}{\xrightleftharpoons}
\newcommand{\xa}{\xrightarrow[]}
\renewcommand{\c}{\color{black}}
\newcommand{\ch}{\color{black}}

\renewcommand{\b}{\beta}
\renewcommand{\d}{\delta}

\newcommand{\e}{\epsilon}
\newcommand{\s}{\sigma}

\begin{document}

\title{How flexibility can enhance catalysis}
\author{Olivier Rivoire}
\affiliation{Center for Interdisciplinary Research in Biology (CIRB), Coll\`ege de France, CNRS, INSERM, Universit\'e PSL, Paris, France.}
\affiliation{Gulliver UMR CNRS 7083, ESPCI, Universit\'e PSL, Paris, France.}

\begin{abstract}
Conformational changes are observed in many enzymes, but their role in catalysis is highly controversial. Here we present a theoretical model that illustrates how rigid catalysts can be fundamentally limited and how a conformational change induced by substrate binding can overcome this limitation, ultimately enabling barrier-free catalysis. The model is deliberately minimal, but the principle it illustrates {\c is} general and consistent with unique features of proteins as well as with previous informal proposals to explain the superiority of enzymes over other classes of catalysts. Implementing the discriminative switch suggested by the model could help overcome limitations currently encountered in the design of artificial catalysts.
\end{abstract}

\maketitle

Enzymes can accelerate chemical reactions to a level currently unmatched by 
artificial catalysts from heterogeneous catalysis~\cite{Greeley.2015}, supramolecular chemistry~\cite{Sanders.1998}, catalytic antibodies~\cite{Tawfik.1994} or computational protein design~\cite{Lovelock.2022}. Could it be that enzymes follow different principles~\cite{Jencks.1975,Kirby.2000,swiegers2008mechanical} or are they simply better~\cite{Knowles.1991}? An often-cited difference between enzymes and other catalysts is that enzymes commonly exhibit conformational changes, including along their catalytic cycle~\cite{Henzler-Wildman.20073id}, while artificial catalysts are generally rigid. Following the principle of transition state stabilization -- the cornerstone of catalysis theory -- indeed leads to the design of rigid catalysts~\cite{Pauling.1946,Tawfik.1994,Lovelock.2022}. On the other hand, the role that flexibility{\c, i.e., degrees of freedom internal to the catalyst,} may play in enzyme catalysis is currently very controversial~\cite{Kamerlin.2010,Kohen.2015,Agarwal.2019}.

To date, the problem has been studied primarily by experimental and computational studies of model enzymes, with general arguments remaining informal~\cite{Jencks.1975,Knowles.1991,Kirby.2000,swiegers2008mechanical,Agarwal.2019,Richard.2019}. Efforts to develop theoretical physics models mostly date from the 1970s and have left the issue unsettled~\cite{Welch.1982}. Inspired by the power of simple physical models to clarify the mechanisms of protein folding~\cite{Pande.1997} and allostery~\cite{rouviere2021emergence}, we use here a minimal model of catalysis to demonstrate in the simplest and clearest terms how catalysis can benefit from {\c a particular form of} flexibility {\c where a switch occurs between conformations of very different energy.}

This approach extends our previous studies of complete catalytic cycles with simple physical models that take into account both geometric and energy constraints~\cite{Rivoire.2020,MunozBasagoiti.2022}. {\c These works showed that flexibility is not necessary for catalysis and could even be detrimental: the best catalysts that were found were rigid, with no internal degree of freedom. As solid surfaces in heterogeneous catalysis, they verify the Sabatier principle~\cite{sabatier1920catalyse,Medford.2015}: they cannot lower the energy barrier of a reaction without limiting the desorption of products. Enzymes, on the other hand, are not subject to this trade-off and can reach a diffusion limit where the only limitation is the rate of encounter with the reactants~\cite{Knowles.1977}. We show here how this is possible with a particular form of flexibility that we call a discriminative switch.}

To explore the design space of catalysts beyond rigid constructs, {\c a general but tractable modeling framework is needed. Our previous models were either limited to one dimension~\cite{Rivoire.2020} or required molecular dynamics simulations~\cite{MunozBasagoiti.2022}. Here} we reformulate the problem with a lattice model amenable to efficient {\c and accurate} calculations. This model recapitulates {\c our previous results and extends them in two aspects. First, it} allows us to identify a limit {\c on the efficiency of rigid catalysts. To this end, we quantify the extent $a$ to which catalysis reduces the activation energy of a reaction (the energy that appears in Arrhenius law) from a value $h_s^+$ in the absence of catalyst to a lesser value $ah_s^+$ with $a<1$ in its presence. We show that $a$ has a non-zero lower bound when the catalyst is rigid. Second, our model allows us to expose a generic principle by which} a conformational switch can overcome this limit and enable barrier-free catalysis{\c, with $a=0$. This principle formalizes a key difference between biological and non-biological catalysts}. 

\begin{figure}[t]
\begin{center}
\includegraphics[width=\linewidth]{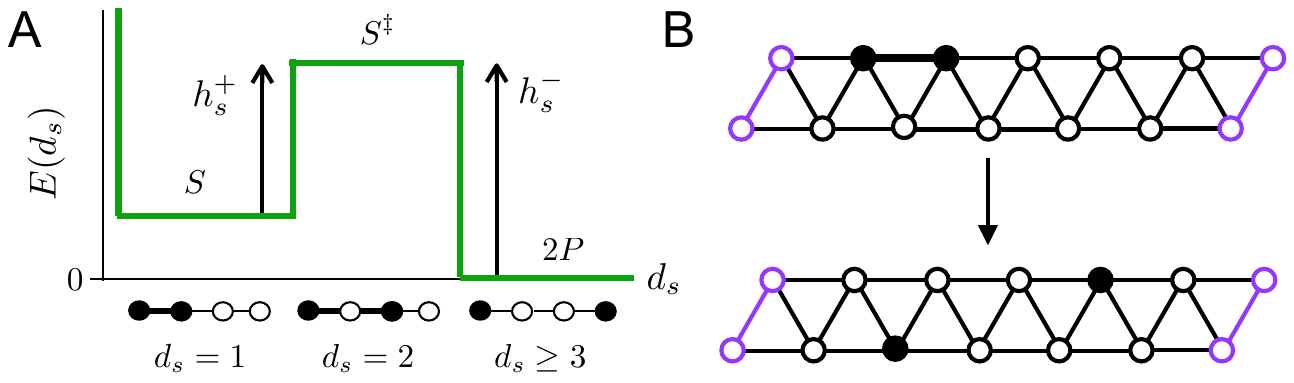}
\caption{{\c Model for the spontaneous reaction --} {\bf A.} Potential of interaction between two particles {\c as a function of their distance $d_s$}, parametrized by the barriers $h_s^\pm$. {\ch No two particles can occupy the same site and therefore $E_s( d_s=0)=\infty$.} {\bf B.} {\c Two particles occupy distinct sites on a} triangular lattice with 12 nodes and periodic boundary conditions along {\c one} direction (repeated purple nodes). {\c The particles initially form a dimer ($d_s=1$, with energy $h_s^--h_s^+$, e.g., top configuration). Each particle can diffuse to a neighboring site with rates given by {\ch Metropolis rule}. The reaction is completed when the particles are free ($d_s\geq 3$, energy 0, e.g., bottom configuration). This requires crossing a transition state ($d_s=2$, energy $h_s^-$) and takes a mean time $T_{S\to 2P}$ that scales as $T_{S\to 2P}\sim e^{h_s^+}$ for large $h_s^+$.}
\label{fig:scheme}}
\end{center} 
\end{figure}

{\it Spontaneous reaction --} As in our previous works~\cite{Rivoire.2020,MunozBasagoiti.2022}, we consider as a spontaneous reaction the dissociation of a dimer into its two constituent monomers but here on a lattice with two particles interacting through a potential $E_s(d_s)$ that is a function of their distance $d_s$ in the lattice {\c (Fig.~\ref{fig:scheme}A)}. The potential excludes two particles from the same site and has two minima, at $d_s=1$ for the dimer and at $d_s\geq 3$ for the free monomers, with an intermediate transition state at $d_s=2$. We parametrize this potential by the forward and backward potential barriers, $h_s^+$ and  $h_s^-$.

{\c The position of particle $i$ is denoted $x_i$ ($i=1,2$). The joint positions $x=(x_1,x_2)$ of the two particles on the lattice define a configuration with associated energy $E(x)=E_s(d_s)$. Each particle can independently hop to a neighboring lattice site to lead to a new configuration $y$. This is taken to occur with a Metropolis rate, $k(x\to y)=k_0\min(1, e^{-(E(y)-E(x))/k_BT})$, where $k_0$ sets the unit of time and $T$ the temperature, which we fix to $k_0=1$ and $k_BT=1$. Other dynamics could be chosen, e.g., Glauber dynamics, the essential feature being that the dynamics is governed by a master equation satisfying the detailed balance, {\ch of the form $\partial_t\pi=Q{\ch ^\top}\pi$ where $\pi(x,t)$ is the probability to be in configuration $x$ at time $t$ and where $Q_{xy}=k(x\to y)$ for $x\neq y$ with $\sum_yQ_{xy}=0$.} Starting from any configuration $x$ where the particles are bound ($d_s=1$), we consider the first time at which the particles are free ($d_s=3$). These sets of initial and final configurations are denoted $S$ and $2P$. Averaging the times over all initial configurations $S$ and over all trajectories ending in $2P$ defines the mean first-passage time $T_{S\to 2P}$, which quantifies the rate of the spontaneous reaction~\cite{ninio1987alternative,MunozBasagoiti.2022}.}

{\c One approach to estimate $T_{S\to 2P}$ is to perform kinetic Monte Carlo simulations~\cite{voter2007introduction}.} For our purposes, as the geometry of the lattice is not critical, we consider a small triangular lattice with $N=12$ sites (Fig.~\ref{fig:scheme}B) {\c for which {\ch we compute $T_{S\to 2P}$ directly by solving} the set of linear equations $\sum_{y\notin 2P}Q_{xy}T_{y\to 2P}=-1$~\cite{redner2001guide,iyer2016first}, from which $T_{S\to 2P}$ is obtained by averaging $T_{x\to 2P}$ over all $x$ in $S$ [see Supplemental Material (SM)].

{\c An analytical solution can also be obtained in the limit of high reaction barriers ($h_s^+\gg1$) when diffusion is negligible compared to barrier crossing and all configurations of same energy are effectively equivalent{\ch ~\cite{pavliotis2008multiscale,zhang2015asymptotic}}. The dynamics then reduces to a three-state Markov chain,
\beq
S\harp[\r_{-1}]{\r_1}S^\ddagger\harp[\r_{-2}]{\r_2}2P
\eeq
where $S$ represents bound configurations ($d_s=1$), $S^\ddagger$ those in the transition state ($d_s=2$) and $2P$ those where the particles are free ($d_s\geq 3$). Assuming instantaneous diffusion at rate 1, the transition rates between these states are simply $\r_1=e^{-h_s^+}$, $\r_{-1}=\r_2=1$ and $\r_{-2}=e^{-h_s^-}$. $T_{S\to 2P}$ is then the solution of just two linear equations which can be solved analytically to yield $T_{S\to 2P}=1/\r_1+1/\r_2+\r_{-1}/(\r_1\r_2)$ (see SM). Given the assumption $h_s^+\gg 1$, we verify Arrhenius law $T_{S\to 2P}\simeq 1/\r_1= e^{h_s^+}$ with $h_s^+$ defining the activation energy for the reaction in the absence of catalyst.}

{\it Rigid catalysis --} {\c A catalyst effectively reduces this activation energy without being modified in the process. Inspired by heterogeneous catalysis where catalysts are solid surfaces, we first consider a catalyst consisting of two binding sites at fixed locations on the lattice (Fig.~\ref{fig:rigid}A). When a particle occupies a binding site, the energy is lowered by $\e_{cs}$, which represents the substrate-catalyst interaction energy (Fig.~\ref{fig:rigid}B). The distance between the binding sites is fixed to $L_c=2$, which is the only value of $L_c$ at which catalysis can occur: if $L_c=1$, binding to both sites stabilizes the dimer and therefore increases the activation energy, while if $L_c\geq 3$, the binding sites cannot impact the transition from $d_s=1$ to $d_s=2$. A geometry with $L_c=2$ is consistent with Pauling principle~\cite{Pauling.1946} which states that a catalyst should be complementary to the transition state, here $d_s=2$.}

{\c By definition, catalysis occurs when the reaction is completed faster in the presence of the binding sites than in their absence. The rate of the catalyzed reaction is estimated by a mean first-passage time denoted $T_{C+S\to C+2P}$, which is defined and obtained as $T_{S\to 2P}$, except that we account for the interaction with the binding sites when computing the energy $E(x)$ of a configuration $x$, and that we restrict the initial and final configurations $C+S$ and $C+2P$ to configurations with no particle at any of the binding sites. We find that catalytic efficiency, defined by $\eta=T_{S\to 2P}/T_{C+S\to C+2P}$ can be $>1$ (the definition of catalysis) with an optimum at an intermediate value of $\e_{cs}$ (Fig.~\ref{fig:rigid}C). 

This result captures Sabatier principle~\cite{sabatier1920catalyse}:  an efficient catalysis must neither bind too weakly nor too strongly to the substrate. This principle is widely observed in heterogeneous catalysis~\cite{Medford.2015} and was also observed in previous off-lattice models where the optimal catalysts similarly consisted of two rigidly held binding sites~\cite{Rivoire.2020,MunozBasagoiti.2022}. Furthermore, we observe that catalysis ($\eta>1$) depends not only on the forward reaction barrier $h_s^+$, but also on the reverse barrier $h_s^-$, and that $\eta\leq e^{h_s^+/2}$ (Fig.~\ref{fig:rigid}E), indicating that the activation energy is at best reduced by a factor $a=1/2$.

\begin{figure}[t]
\begin{center}
\includegraphics[width=.9\linewidth]{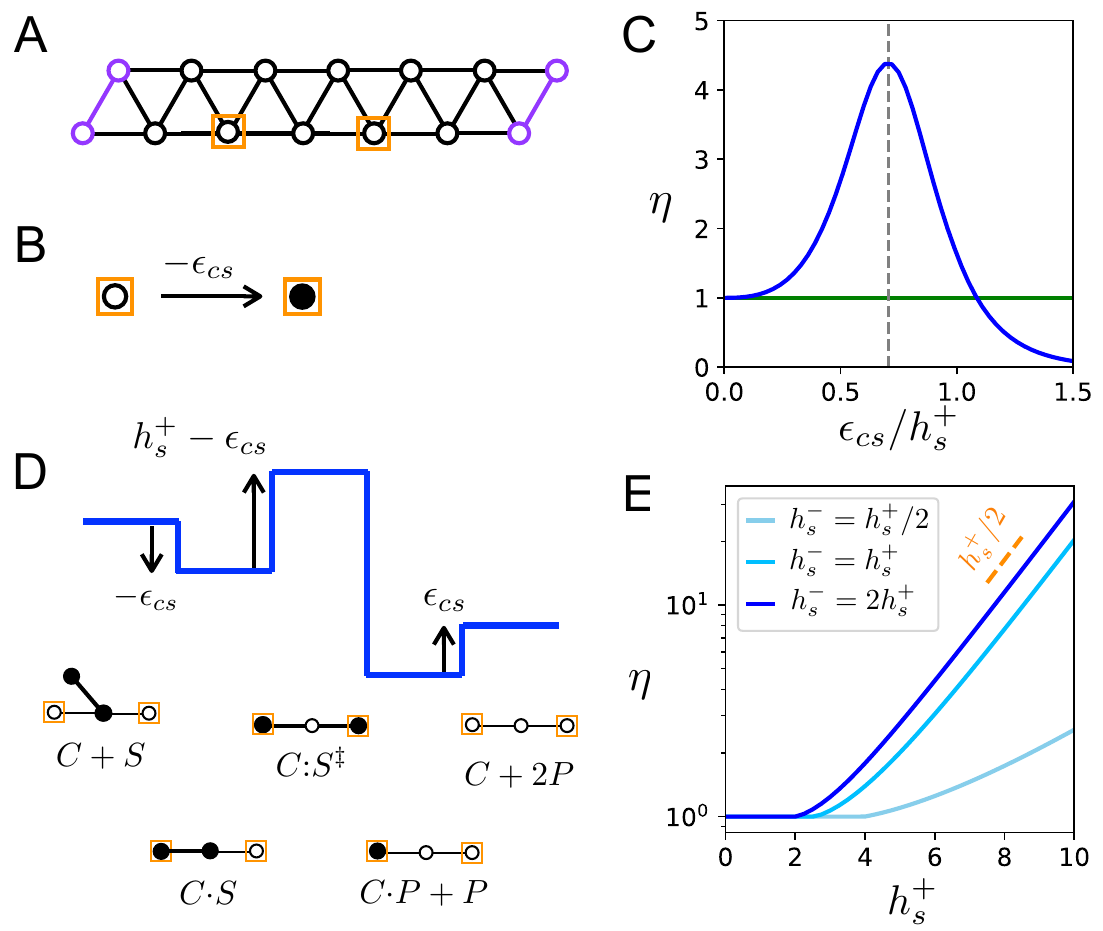}
\caption{{\c Model for rigid catalysis -- {\bf A.} A catalyst consists in two binding sites at fixed locations on the lattice (orange squares). {\bf B.} When a particle occupies one of these sites, the energy is decreased by $\e_{cs}$. {\bf C.}~Catalytic efficiency $\eta=T_{S\to 2P}/T_{C+S\to C+2P}$ comparing the mean reaction time in this model, $T_{C+S\to C+2P}$, with the mean time $T_{S\to 2P}$ for the spontaneous reaction (Fig.~\ref{fig:scheme}) as a function of the binding strength $\e_{cs}$ ($h_s^+=6$ and $h_s^-=12$). This graph is comparable to so-called volcano plots in heterogeneous catalysis~\cite{Bligaard.2004}. {\bf D.}~Energy landscape illustrating how a rigid catalyst replaces the single barrier $h_s^+$ (Fig.~\ref{fig:scheme}A) by two smaller barriers $h_s^+-\e_{cs}$ and $\e_{cs}$. The occupation of the binding sites in each configuration is represented at the bottom.} {\bf E.} Catalytic efficiency of optimal rigid catalysts as a function of the forward barrier $h_s^+$ for different reverse barrier $h_s^-$, showing that catalysis involves a threshold beyond which the efficiency scales exponentially.
\label{fig:rigid}}
\end{center} 
\end{figure}

These observations are rationalized by studying analytically the limit of high reaction barriers ($h_s^+\gg 1$). In this limit, the dynamics can again be reduced to a Markov process with only few states, this time five,
\beq
C+S\harp[\r_{-1}]{\r_1}C\.S\harp[\r_{-2}]{\r_2}C\:S^\ddagger\harp[\r_{-3}]{\r_3}C\.P+P\harp[\r_{-4}]{\r_4}C+2P
\eeq
where $C+S$ represents configurations with a dimer ($d_s=1$) occupying none of the binding sites, $C\.S$ those with a dimer occupying one binding site, $C\:S^\ddagger$ those where the two binding sites are occupied (and therefore $d_s=2$), $C\.P+P$ those where the particles are unbound ($d_s\geq 3$) but one occupies a binding site and $C+2P$ those where the particles are unbound and none occupies a binding site. 

Given the correspondence between energies and rates, the states and rates can be represented by an energy landscape (Fig.~\ref{fig:rigid}D). This representation illustrates how catalysis by two rigidly held binding sites works: it replaces the single energy barrier $h_s^+$ of the spontaneous reaction (Fig.~\ref{fig:scheme}A) by two smaller energy barriers (Fig.~\ref{fig:rigid}C), a barrier $h_s^+-\e_{cs}$ from $C\.S$ to $C\:S^\ddagger$ and a barrier $\e_{cs}$ from $C\.P+P$ to $C+2P$. Increasing $\e_{cs}$ decreases the first barrier but increases the second, which is the trade-off known as Sabatier principle~\cite{sabatier1920catalyse} and the reason for the non-trivial optimum in Fig.~\ref{fig:rigid}B. 

In the limit $h_s^+\gg 1$, the dynamics is controlled by the highest barrier and the optimum is therefore when the two barriers are the same, which gives $\e_{cs}=h_s^+/2$. This explains why the activation energy can be lowered by a factor $a=1/2$ at best. This result is consistent with the one obtained in an off-lattice model, where $a\simeq 0.56$ at best~\cite{MunozBasagoiti.2022}. For this optimum to be reached, the back reaction $C\.P+P\to C\:S^\ddagger$ must be negligible, or time is spent recrossing the barriers. This explains the role played by the reverse barrier $h_s^-$ (Fig.~\ref{fig:rigid}E). These conclusions are verified by analytical calculations (see SM) showing that the factor $a$ by which a catalyst reduces the activation energy satisfies
\beq\label{eq:abound}
a\geq \frac{1}{2}+\max\left(0,\frac{h_s^+-h_s^-}{2 h_s^+}\right),
\eeq
which implies $a\geq 1/2$, {\ch with $a=1/2$} reachable only if $h_s^->h_s^+$. This analytical result is obtained in the limit $h_s^+\gg 1$ but {\ch analyzing the lattice model shows that it} also provides an upper bound on the catalytic efficiency $\eta$ for large but finite values of $h_s^+$ (Fig.~S1).}

{\c To try to go beyond the limit of Eq.~\eqref{eq:abound}, several extensions of the model may be contemplated. For instance, we may consider the two binding sites to have different binding energies $\e_{cs}^1\neq \e_{cs}^2$. However, the site with highest energy that most lowers $h_s^+$ is also inevitably the one that most limits release, implying a symmetric optimum with $\e_{cs}^1= \e_{cs}^2$ (SM and Fig.~S2). Alternatively, we may consider relaxing the assumption that the binding sites are fixed. For instance, we may assume them to fluctuate between a conformation with $L_c=2$ and another with $L_c=1$, possibly with an energy difference $\e_c$. This is the type of flexibility ``along the reaction coordinate'' considered in previous models~\cite{Rivoire.2020,MunozBasagoiti.2022} and we verify again here that it is detrimental to catalysis (SM and Fig.~S3). This is simply explained: only when $L_c=2$ is the energy barrier from $d_s=1$ and $d_s=2$ effectively lowered.}

{\c {\it Catalysis with a discriminative switch} --  The limitation expressed by Eq.~\eqref{eq:abound} is in sharp contrast with the evidence that some enzymes can effectively totally annihilate activation barriers~\cite{Knowles.1977}, which in our model corresponds to $a=0$. This indicates that breaking the trade-offs of Sabatier principle is possible but by a mechanism that must differ from those considered previously.}

{\c We demonstrate here such a mechanism, which we call a discriminative switch. In this design, the catalyst {\ch can be in two} states, $C_0$ and $C_1$, with the latter having a larger energy $\e_c$. These internal states are represented in our model by a third particle confined to two additional lattice sites while the two binding sites are kept at the same fixed locations (Fig.~\ref{fig:2state}A). The internal states are coupled to the reaction without compromising the geometry or rigidity of the binding sites themselves. This is achieved by interaction energies that depend on the internal state of the catalyst: when in state $C_0$, the binding energy is as before $\e_{cs}$ but when in state $C_1$ an additional contribution brings it to $\e_{cs}+\d_{cs}$ (Fig.~\ref{fig:2state}B). In our model, $\d_{cs}$ may be thought as arising from the interaction with the third particle which is brought closer to the binding sites in state $C_1$. In enzymes, this could correspond to additional interactions arising when a surface loop that is flexible in an open state $C_0$ comes to surround the substrate in a closed state $C_1$, with an associated entropy loss $\e_c$. 

The key is then to make the following four choices that guarantee that no step involves a positive energy barrier: (1) $\e_c=\d_{cs}$ so that the transitions between states where a single site is bound, $C_0\.S\harp{} C_1\.S$ and $C_0\.P\harp{} C_1\.P$, are barrier-less, (2) $\d_{cs}\geq h_s^+$ so that accessing $S^\ddagger$ through $C_1\.S\to C_1\:S^\ddagger$ involves a reduction of energy, (3) $\d_{cs}\leq h_s^-$ so that no up-hill barrier is introduced for $C_1\:S^\ddagger\to C_1\.P+P$, and (4) $\e_{cs}=0$ so that release $C_0\.P+P\harp{} C_0+2P$ is barrier-less. With such parameters, i.e., with
\beq\label{eq:conditions}
\e_{cs}=0\quad{\rm and}\quad h_s^+\leq \d_{cs}=\e_c\leq h_s^-
\eeq
a path from $C_0+S$ to $C_0+2P$ is defined along which the energy of the system does not increase at any step (Fig.~\ref{fig:2state}D), provided $h_s^+<h_s^-$.}

\begin{figure}[t]
\begin{center}
\includegraphics[width=.95\linewidth]{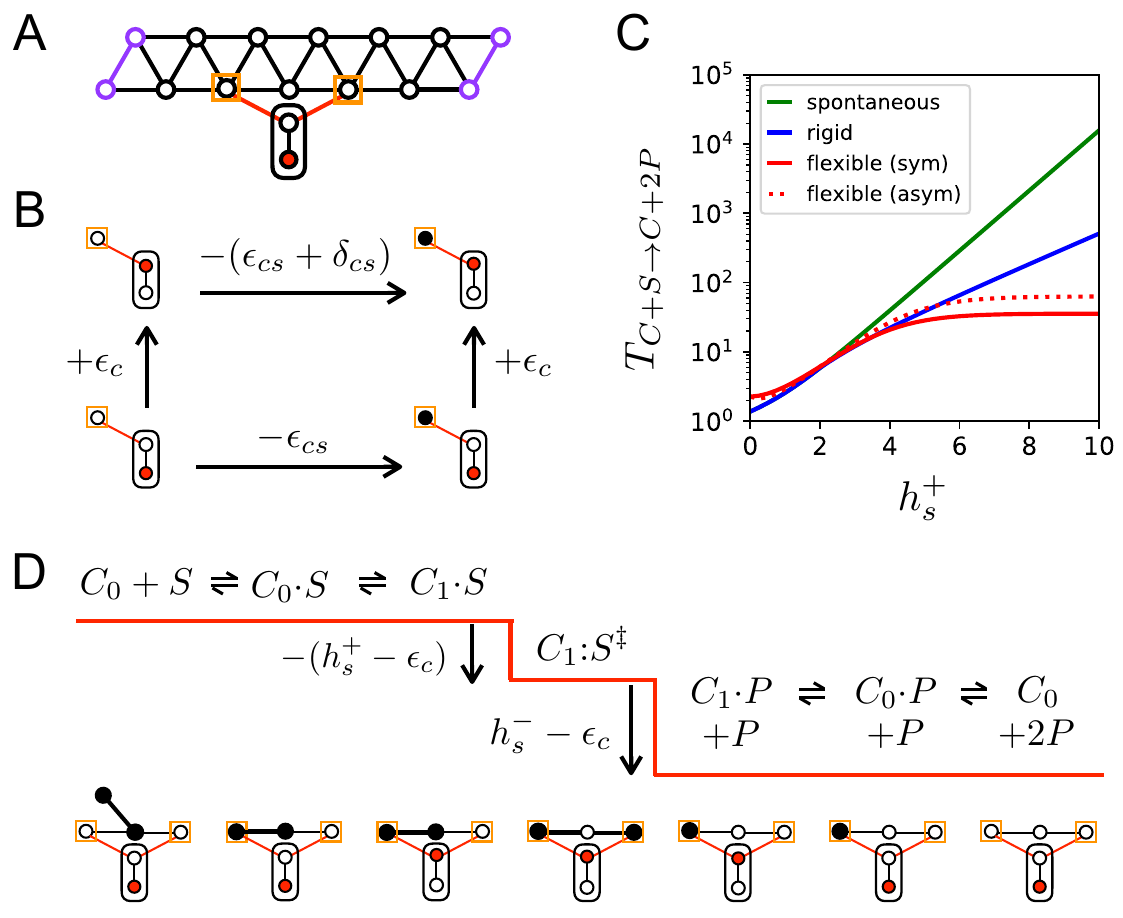}
\caption{{\c Model for catalysis with a discriminative switch -- 
{\bf A.}~We add a degree of freedom in the form of a red particle that can take two positions, down ($C_0$) or up ($C_1$). {\bf B.}~Switching from $C_0$ to $C_1$ involves an energy cost $\e_c$ and the interaction energy at each binding site is $\e_{cs}$ in $C_0$ but $\e_{cs}+\d_{cs}$ in $C_1$. {\bf C.}} Mean time of completion of the reaction in the absence of catalyst (green line), in the presence of an optimal rigid catalyst (blue line), a flexible symmetric catalyst with $\e_{cs}=0$ and $\e_c=\d_{cs}=(h_s^++h_s^-)/2$ (full red line) or a flexible asymmetric catalyst with $\e^1_{cs}=\e^2_{cs}=0$, $\e_c=\d^1_{cs}=4h_s^+$ and $\d^2_{cs}=1.5h_s^+$ (dotted red line). {\c For large $h_s^+$, $T_{C+S\to C+2P}$ scales as} $e^{a h_s^+}$ where $a=1$ without catalyst, $a\geq 1/2$ with rigid catalysts but $a=0$ with catalysts having a discriminative switch {\ch (with $h_s^-$ taken to be $h_s^-=2h_s^+$). Standard deviations can also be computed to show that the distributions of first-passage times are increasingly distinct as  $h_s^+$ increases (Fig.~S5).} {\c {\bf D.}~When the conditions given in Eq.~\eqref{eq:conditions} are satisfied, the energy along the path from $C_0+S$ to $C_0+2P$ does not increase at any step. Catalysis is then barrier-less. Examples of configurations along this path are illustrated at the bottom.}\label{fig:2state}}
\end{center} 
\end{figure}

{\c These arguments are borne out by numerical and analytical calculations that follow the same principles as previously, with the internal state of the catalyst treated as a third particle. This is illustrated in Fig.~\ref{fig:2state}C where $T_{C+S\to C+2P}$ in the presence of the catalyst of Fig.~\ref{fig:2state}A is found to plateau as $h_s^+$ increases, consistent with $T_{C+S\to C+2P}\sim e^{ah_s^+}$ with $a=0$ (barrier-less catalysis).}

{\c A variant of the model can also be defined where the two binding sites are not equivalent but have different parameters $\e_{cs}^1,\d_{cs}^1$ and $\e_{cs}^2,\d_{cs}^2$. This leads to different conditions for barrier-less catalysis (see SM)
\bea\label{eq:asym}
&\e^1_{cs}=0,\quad \e^2_{cs}\leq 0,\quad  \d^2_{cs}\leq\d^1_{cs}=\e_c,\nonumber\\
&{\rm and}\quad h_s^+\leq \e^2_{cs}+\d^2_{cs}\leq h_s^-
\eea
(or the same conditions with the roles of sites 1 and 2 reversed). Here, $\e_{cs}^1=0$ and $\d_{cs}^1=\e_c$ guarantee barrier-less transitions $C_0+S\harp{}C_0\.S\harp{}C_1\.S$ and $C_1\.P+P\harp{}C_0\.P+P\harp{}C_0+2P$ (with binding at site 1) while $h_s^+\leq \e_{cs}^2+\d_{cs}^2$ guarantees that $C_1\.S\to C_1\:S^\ddagger$ is down-hill. The additional constraint $\d^2_{cs}\leq\e_c$ is necessary to prevent a particle to be stuck at binding site 2. It is indeed generally not sufficient to have a barrier-less path from reactant to products for barrier-less catalysis to occur as alternative paths may be present that lead to kinetic traps. This asymmetric design also achieves barrier-less catalysis {\ch (Fig.~\ref{fig:2state}C) although with a lower catalytic efficiency (Fig.~S4),} but it is more comparable to enzymes whose substrates are typically asymmetric and where a distinction is usually made between a ``binding site'' (site 1) and an ``active site'' (site 2).

{\c {\it Conclusion --} Based on the formulation and solution of an elementary model of catalysis, we have illustrated how a particular form of flexibility involving a two-state switch can overcome the limitations of rigid catalysis and effectively enable barrier-less catalysis, where the activation energy of the spontaneous reaction is totally annihilated. The expression for the bound on rigid catalysis given by Eq.~\eqref{eq:abound} is specific to our model but reflects a fundamental trade-off widely observed in heterogeneous catalysis where it is known as Sabatier principle~\cite{Medford.2015}. The mechanism that we demonstrated, by which this bound can be overcome with a switch involving a compensation between two large (free) energies ($\e_c$ and $\d_{cs}$) is generic and directly echoes the proposal that biological catalysts differ from non-biological catalysts by their use of an ``intrinsic binding energy''~\cite{Jencks.1975}. This concept has been illustrated in enzymes~\cite{Richard.2019} and ribozymes~\cite{hertel1997use} but its links to catalytic rate enhancements, product release and conformational changes have never been fully explained, as reflected by the controversies over the role of flexibility in enzyme catalysis~\cite{Kamerlin.2010,Kohen.2015,Agarwal.2019} and the absence of this concept in reflections to overcome Sabatier principle in heterogeneous catalysis~\cite{perez2019strategies}. 
Our model clearly exposes these different links. Finally, from a physics standpoint, our modeling approach and our results are of interest for studying the many physical phenomena involving a coupling between a chemical reaction and a conformational change.}

\begin{acknowledgements}
I am grateful to C. Nizak,  M.~Mu\~noz Basagoiti, Y. Sakref, Z. Zeravcic, and I. Junier for discussions, and to ANR-21-CE45-0033 for funding.
\end{acknowledgements}

\clearpage
\newpage
\appendix

\makeatletter
\makeatletter \renewcommand{\fnum@figure}
{\figurename~S\thefigure}
\makeatother
\setcounter{figure}{0}

\renewcommand\theequation{S\arabic{equation}}
\setcounter{equation}{0}

\onecolumngrid

\vspace{1cm}
\centerline{\bf \Large Supplemental Material}
\vspace{2cm}

\twocolumngrid

\section*{Model definition}

{\c For the most general model of flexible catalysis with a conformational switch, from which the models for the spontaneous reaction and for rigid catalysis are obtained as limit cases, a} configuration $x=(x_1,x_2,\s_c)$ of the system consists of the locations $x_1,x_2$ of the two particles on the lattice together with the state $\s_c$ of the catalyst, which may be open ($\s_c=0$) or closed ($\s_c=1$).  {\c In this representation, $x_i=1,\dots,N$ for each of two particles ($i=1,2$) where $N$ is the total number of lattice sites ($N=12$ in the results presented in Figs~2-3). Given that the two particles {\ch are indistinguishable and} cannot occupy the same site, the total number of configurations is $N(N-1)$.} We denote by $d_s=d(x_1,x_2)$ the distance between the particles, defined by the length of the shortest connecting path on the lattice, and by $\s_k=1$ the occupancy of binding site $k$, with $\s_k=0$ indicating that it is vacant ($k=1,2$ {\c when considering two binding sites}). In terms of these variables, the energy of configuration $x$ is
\beq
E(x)=E_s(d_s)+E_c(\s_c)-\sum_{k=1,2}E^k_{cs}(\s_c)\s_k
\eeq
where
\beq
E_s(d_s) = \left\{
    \begin{array}{ll}
        +\infty & \mbox{if } d_s=0 \\
       h_s^--h_s^+ & \mbox{if } d_s=1\\
       h_s^- & \mbox{if } d_s=2\\
       0 & \mbox{if } d_s\geq 3,\\
    \end{array}
\right.
\eeq
\beq
E_c(\s_c)=\e_c\s_c
\eeq
and
\beq
E^k_{cs}(\s_c)=\e_{cs}^k+\d_{cs}^k\s_c
\eeq
{\c The spontaneous reaction corresponds to $\e_{cs}^1=\e_{cs}^2=0$ (Fig.~1) and rigid catalysts to $\e_c=\infty$ and $\e_{cs}^1=\e_{cs}^2=\e_{cs}$ in the symmetric case (Fig.~2). We also introduce below an extension of the model to mobile binding sites.}

The system can transition from a configuration $x=(x_1,x_2,\s_c)$ to any of the configurations $y$ given by $(y_1,x_2,\s_c)$, $(x_1,y_2,\s_c)$, $(x_1,x_2,1-\s_c)$ where $d(x_1,y_1)=1$ and $d(x_2,y_2)=1$. These transitions occur with Metropolis rates given by
\beq
k(x\to y)={\rm min}(1,e^{-(E(y)-E(x))}).
\eeq
This defines a master equation with detailed balance in the form of a continuous-time discrete Markov process,
\beq
\partial_t\pi(x,t)=\sum_{y\neq x}\left(\pi(y,t)k(y\to x)-\pi(x,t)k(x\to y)\right)
\eeq
where $\pi(x,t)$ is the probability to be in configuration $x$ at time $t$. {\c This {\ch forward} master equation can be written in matrix form as $\partial_t\pi(t)=Q^{\ch \top}\pi(t)$ where $\pi(t)$ is a $N(N-1)$-dimensional vector with components $\pi(x,t)$ and where {\ch $Q^{\top}$ is the transpose of the $(N(N-1))\times (N(N-1))$ dimensional matrix $Q$ that defines the generator of the continuous time Markov chain, whose components are}
\bea\label{eq:Q}
Q_{xy}=
\left\{
    \begin{array}{ll}
        k(x\to y) & \mbox{if } x\neq y \\
       -\sum_{y\neq x}k(y\to x)& \mbox{if } x=y.\\
    \end{array}
\right.
\eea
}

{\ch 
\section*{Limit of fast diffusion}

In the limit of large barriers $h_s^+\to\infty$, diffusion becomes negligible and the dynamics can be approximated by a Markov process with a fewer number of states. Formally, the procedure is known as ``averaging'' and relies on a clustering of the configurations into subsets such that intra-transition rates within the subsets are negligible compared to inter-transition rates between subsets~
\cite{pavliotis2008multiscale,zhang2015asymptotic}. In our model, the different subsets are defined by the bonds that are formed between the particles and the catalyst. They gather configurations of same energy that are connected through intra-transitions with rates $k(x\to y)=1$.
}

\section*{\ch Moments of first-passage time}

For the spontaneous reaction, we define $S$ as the set of configurations with $d_s=1$ and $2P$ as the set of configurations with $d_s\geq 3$. With $T_{x\to 2P}$ denoting the mean first-passage time from $x\in S$ to $2P$, we define the global mean first-passage time $T_{S\to 2P}$ by
\beq
T_{S\to 2P}=\frac{1}{|S|}\sum_{x\in S}T_{x\to 2P}
\eeq
where $|S|$ is the size of set $S$.

{\c For reactions in presence of a rigid catalyst}, we define $C+S$ as the set of configurations with $d_s=1$ and $\s_1=\s_2=0$, and $C+2P$ as the set of configurations with $d_s\geq 3$ and $\s_1=\s_2=0$, {\c where the constraint $\s_1=\s_2=0$ enforces that the binding sites are initially and finally free}. We define $T_{C+S\to C+2P}$ as
\beq
T_{C+S\to C+2P}=\frac{1}{|C+S|}\sum_{x\in C+S}T_{x\to C+2P}.
\eeq

{\c For reactions in presence of a two-state catalyst we further impose the state of the catalyst to be in the same conformational state of lowest energy ($C_0$) in the initial and final configurations.}

We compute mean first-passage times {\c numerically} by linear algebra, {\ch using the fact that the distribution of first passage times $f_B(t,x)$ from a configuration $x\notin B$ to a set $B$ follows the backward Master equation $\partial_t f_B(t,x)=\sum_y Q_{xy}f_B(t,y)$ with the matrix $Q$ defined in Eq.~\eqref{eq:Q}~\cite{iyer2016first}. This relationship implies that the mean first-passage time $T_{x\to B}=\int_0^\infty \tau f_B(\tau,x)d\tau$ is solution of the equations $\sum_{y\notin B}Q_{xy} T_{y\to B}=-1$.} Introducing  $\tilde Q$ defined over configurations not in $B$ by $\tilde Q_{xy}=Q_{xy}$, this corresponds to solving the matrix equation $\tilde Q T_{\cdot\to B}=-U$ where $U$ is a vector whose components are all one. 

{\ch This approach generalizes to the computation of the $n$-th moment of the first-passage time, $T^{(n)}_{x\to B}=\int_0^\infty \tau^n f_B(\tau,x)d\tau$, which can be obtained by solving $\tilde Q^n T^{(n)}_{\cdot\to B}=(-1)^nnU$, with the mean first passage time $T_{x\to B}= T^{(1)}_{x\to B}$ corresponding to the particular case $n=1$. The standard deviations of first passage times represented in Fig.~S\ref{fig:Snew} are computed from the first and second moments as $(T^{(2)}-(T^{(1)})^2)^{1/2}$.}

{\c \section*{Mean first-passage times for 1d Markov chains}

The previous formalism can be applied to obtain analytical expressions when considering one-dimensional Markov chains.

When the chain is of length 2, of the form
\beq
A_1\harp[\r_{-1}]{\r_{1}} A_2\harp[\r_{-2}]{\r_{2}} A_3
\eeq
where $\r_i$ represents the forward rate from $A_i$ to $A_{i+1}$ and where $\r_{-i}$ the backward rate from $A_{i+1}$ to $A_i$, the matrix $\tilde Q$ is given by
\beq
\tilde Q=
\begin{bmatrix}
-\r_1 & \r_1\\
\r_{-1} & -(\r_{-1}+\r_2)
\end{bmatrix}
\eeq
and applying the formula $T_{A_1\to A_3}=-(\tilde Q^{-1} U)_1$ with $U=[1,1]$ leads to
\beq\label{eq:1d2}
T_{A_1\to A_3}=\frac{1}{\r_1}+\frac{1}{\r_2}+\frac{\r_{-1}}{\r_1\r_2}.
\eeq

Similarly, for a Markov chain of length 3 given by
\beq
A_1\harp[\r_{-1}]{\r_{1}} A_2\harp[\r_{-2}]{\r_{2}} A_3\harp[\r_{-3}]{\r_{3}} A_4
\eeq
we obtain
\beq\label{eq:1d3}
T_{A_1\to A_4}=\frac{1}{\r_1}+\frac{1}{\r_2}+\frac{1}{\r_3}+\frac{\r_{-1}}{\r_1\r_2}+\frac{\r_{-2}}{\r_2\r_3}+\frac{\r_{-1}\r_{-2}}{\r_1\r_2\r_3}.
\eeq
}

\section*{Spontaneous reaction}

In the {\c limit of large reaction barrier $h_s^+\gg 1$}, we can ignore the contribution of diffusional processes and estimate $T_{S\to 2P}$ as the mean first-passage time from $S$ to $2P$ for the Markov chain 
\beq
S\harp[\r_{-1}]{\r_{1}} S^\ddagger \xa{\r_2} 2P
\eeq
with, since $h_s^+>0$ and $h_s^->0$,
\beq
\r_{1}\sim e^{-h_s^+},\quad \r_{-1}\sim 1,\quad \r_2\sim1
\eeq
as the pre-factors play no role in the {\c $h_s^+\to\infty$} limit. {\c Applying Eq.~\eqref{eq:1d2},}
\beq
T_{S\to 2P}=\frac{1}{\r_1}+\frac{1}{\r_2}+\frac{\r_{-1}}{\r_1\r_2}.
\eeq
which leads to {\c $T_{S\to 2P}\sim e^{h_s^+}$ or, more formally,
\beq
\lim_{h_s^+\to\infty}\frac{1}{h_s^+}\ln T_{S\to 2P}=1.
\eeq
}

\section*{{\c Catalysis with symmetric binding sites}}

We analyze here {\c the model of Fig.~2 with two fixed binding sites with same interaction energy $\e_{cs}$}. As for the spontaneous reaction, the problem reduces to a Markov chain {\c in the limit where diffusion is negligible ($h_s^+\gg 1$ and $\e_{cs}\gg 1$)}. Here, the relevant states are, in addition to $C+S$ and $C+2P$, state $C\.S$ defined by $d_s=1$ and $\s_1+\s_2=1$, state $C\:S$ defined by $d_s=2$ and $\s_1+\s_2=2$ and state $C\.P+P$ defined by $d_s\geq 3$ and $\s_1+\s_2=1$.

To obtain a lower bound on $T_{C+S\to C+2P}$, we can start from $C\.S$ and consider
\beq
C\.S\harp[\r_{-2}]{\r_{2}} C\:S \harp[\r_{-3}]{\r_{3}} C\.P+P\xa{\r_4}C+2P
\eeq
to which is associated {\c a mean first-passage time given by Eq.~\eqref{eq:1d3} (with a shift in the indices since we consider $C\.S$ and not $C+S$ as initial state),}
\beq
T_{C\.S\to C+2P}=\frac{1}{\r_2}+\frac{1}{\r_3}+\frac{1}{\r_4}+\frac{\r_{-2}}{\r_2\r_3}+\frac{\r_{-3}}{\r_3\r_4}+\frac{\r_{-2}\r_{-3}}{\r_2\r_3\r_4}.
\eeq

If $\e_{cs}>h_s^+$, then $1/\r_4\sim e^{\e_{cs}}$ cannot be lower than $T_{S\to 2P}$. We therefore consider $\e_{cs}<h_s^+$. Next, if $ \e_{cs}>h_s^-$, $\r_{-2}\r_{-3}/(\r_2\r_3\r_4)\sim e^{h_s^+-h_s^-+\e_{cs}}$ cannot be lower than $T_{S\to 2P}$. We therefore also consider $\e_{cs}<h_s^-$. Under these conditions we have
\bea
&\r_{2}&\sim e^{-(h_s^+-\e_{cs})},\quad \r_{-2}\sim 1,\quad \r_{3}\sim 1\nonumber\\
&\r_{-3}&\sim e^{-(h_s^--\e_{cs})},\quad \r_4\sim e^{-\b\e_{cs}}.
\eea
It follows that {\c 
\beq
T_{C\.S\to C+2P}\sim e^{h_c^+}
\eeq
with}
\beq\label{eq:dgc}
h_c^+=\max(h_s^+-\e_{cs},\e_{cs}, 2\e_{cs}-h_s^-,h_s^+-h_s^-+\e_{cs}).
\eeq

\begin{figure}[t]
\begin{center}
\includegraphics[width=.6\linewidth]{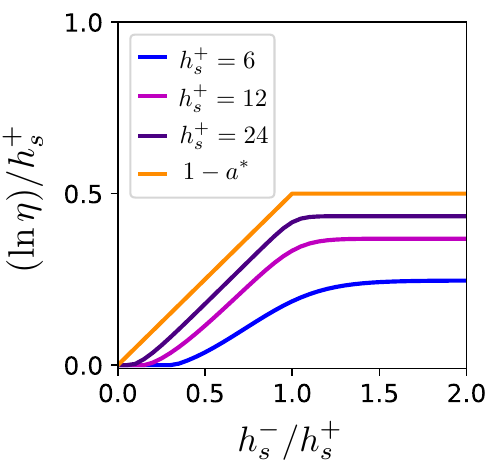}
\caption{\c Bound on catalytic efficiency for finite barriers -- Optimal catalytic efficiency $\eta=T_{S\to 2P}/T_{C+S\to C+2P}$ for a catalyst with two rigidly held binding sites (model of Fig.~2) as a function of the reverse barrier $h_s^-$ for different values of the forward barrier $h_s^+$, showing that $\eta\leq e^{(1-a^*)h_s^+}$ with $a^*$ given by Eq.~\eqref{eq:astar}. To obtain optimal catalytic efficiencies, the interaction energy $\e_{cs}$ is optimized numerically for each value of $(h_s^-,h_s^+)$. \label{fig:S1}}
\end{center} 
\end{figure}

Optimal rigid catalysts have activation energy $\min_{\e_{cs}}h_c^+(\e_{cs})$. In the limit $h_s^-{\c/h_s^+}\to\infty$ of irreversible reactions where $h_c^+(\e_{cs})=\max(h_s^+-\e_{cs}, \e_{cs})$ the optimum over $\e_{cs}$ is obtained for $h_s^+-\e_{cs}=\e_{cs}$, leading to $\e^*_{cs}=h_s^+/2$ and $h_c^+(\e_{cs}^*)=h_s^+/2$. This expression is valid as long as no other term in Eq.~\eqref{eq:dgc} exceeds $h_s^+/2$ when considering $\e_{cs}=\e_{cs}^*$. The largest value of $h_s^-$ at which this ceases to be the case is $h_s^-=h_s^+$ due to the term $h_s^+-h_s^-+\e_{cs}$. For $h_s^-<h_s^+$, this term dominates over $\e_{cs}$ and the optimum is obtained when $h_s^+-\e_{cs}=h_s^+-h_s^-+\e_{cs}$, leading to $\e_{cs}^*=h_s^-/2$ and $h_c^+(\e_{cs}^*)=h_s^+-h_s^-/2$. Finally, $a^*=\min_{\e_{cs}}h_c^+(\e_{cs})/h_s^+$ {\c is given by
\beq\label{eq:astar}
a^*=\frac{1}{2}+\max\left(0,\frac{h_s^+-h_s^-}{2 h_s^+}\right),
\eeq
which is the right-hand side of Eq.~(1) in the main text.

For finite $h_s^+$, we find numerically that catalytic efficiency defined by $\eta=T_{S\to 2P}/T_{C+S\to C+2P}$ verifies $\eta\leq e^{(1-a^*)h_s^+}$, i.e., the same bound applies (Fig.~S\ref{fig:S1}).}

{\c
\section*{Catalysis with asymmetric binding sites}

Here we consider an extension of the model of Fig.~2 where the two binding sites can have different binding energies $\e_{cs}^1$ and $\e_{cs}^2$. Numerically, we observe that the catalytic efficiency $\eta=T_{S\to 2P}/T_{C+S\to C+2P}$ is symmetric in $(\e_{cs}^1,\e_{cs}^2)$  with an optimum when $\e_{cs}^1=\e_{cs}^2$ (Fig.~S\ref{fig:S2}) 

\begin{figure}[t]
\begin{center}
\includegraphics[width=.95\linewidth]{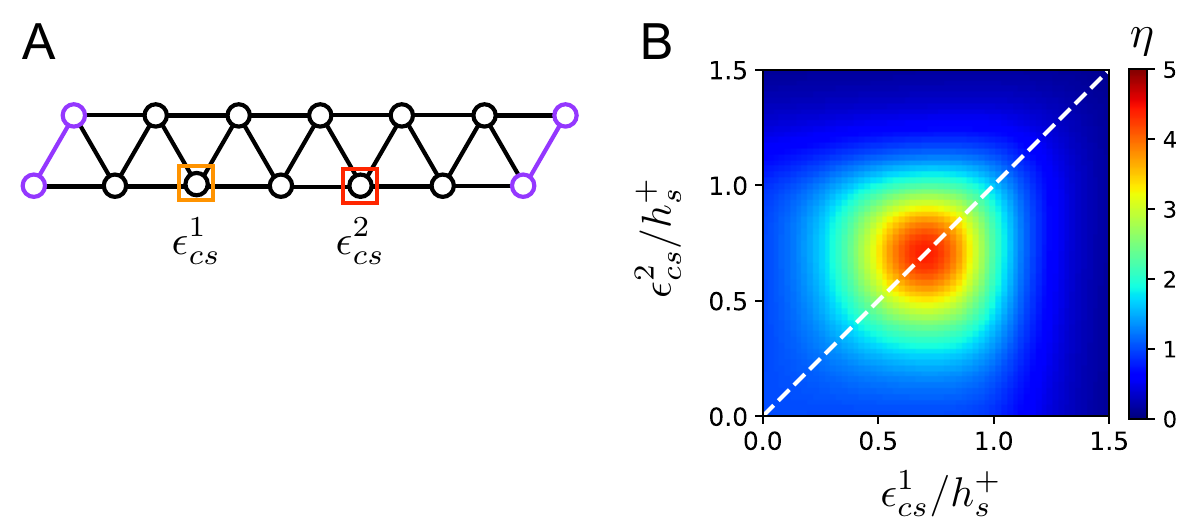}
\caption{\c Catalysis with two rigidly held asymmetric binding sites -- {\bf A.} The model of Fig.~2 is extended to the case where the two binding sites have different binding energies, $\e_{cs}^1$ and $\e_{cs}^2$. {\bf B.}~Catalytic efficiency $\eta=T_{S\to 2P}/T_{C+S\to C+2P}$ as a function of the two binding energies $\e_{cs}^1$ and $\e_{cs}^2$ for $h_s^+=6$ and $h_s^-=12$, showing a symmetry with an optimum when $\e_{cs}^1=\e_{cs}^2$. \label{fig:S2}}
\end{center} 
\end{figure}

We can understand why a symmetric design is optimal by examining the limit $h_s^+,\e_{cs}\gg 1$ where the dynamics reduces to a Markov chain with few states, here of the form
\beq
\begin{tikzcd}[cramped,column sep=tiny,row sep=tiny]
      & C\.S_1 \arrow[dr]\arrow[dl,shift right]  &  		& C\.P_1+P\arrow[dr,shift left]\arrow[dl,shift right]  & 	\\
C+S \arrow[ur]\arrow[dr]  &    & C\:S^\ddagger\arrow[ur]\arrow[dr]\arrow[ul,shift left]\arrow[dr]\arrow[dl,shift right] & 			& C+2P\arrow[ul]\arrow[dl,shift left] \\
					& C\.S_2\arrow[ur]\arrow[ul,shift left]   &						      &	C\.P_2+P\arrow[ur]\arrow[ul,shift left]  & \nonumber
\end{tikzcd}
\eeq
where $C\.S_1$ or $C\.P_1+P$ refers to configuration where binding site 1 is occupied and $C\.S_2$ or $C\.P_2+P$ where binding site 2 is occupied. For simplicity, consider the case $h_s^-\to\infty$ (generally  the most favorable to catalysis) so that there is no possible return to $C\:S^\ddagger$ once a transition is made to either $C\.P_1+P$ or $C\.P_2+P$. As the system may end up in any of these states, the time from $C\:S^\ddagger$ to $C+2P$ is dominated by the largest barrier along the paths $C\.P_1+P\to C+2P$ and $C\.P_2+P\to C+2P$, that is, $\max(\e_{cs}^1,\e_{cs}^2)$. We therefore have 
$T_{C+S\to C+2P}\sim e^{h_c^+}$ with
\beq
h_c^+\geq\max(\min(h_s^+-\e^1_{cs},h_s^+-\e_{cs}^2),\max(\e^1_{cs},\e^2_{cs})).
\eeq
If we assume without loss of generality that $\e^1_{cs}\geq \e^2_{cs}$, this gives $h_c^+\geq\max(h_s^+-\e_{cs}^1,\e^1_{cs})$, the exact same trade-off as in the symmetric case. The optimum is achieved when $\e_{cs}^1=h_s^+/2$, implying $a\geq 1/2$: an asymmetric design cannot yield more efficient catalysis than a symmetric design with $\e_{cs}^1=\e_{cs}^2$. Physically, the interpretation is simple: if one binding site has a larger binding energy, this energy can be used to accelerate the access to the transition state but the same site will also be the one most limiting release, with eventually exactly the same trade-off as in the symmetric case.
}

{\c 
\section*{Catalysis with mobile binding sites}

\begin{figure}[t]
\begin{center}
\includegraphics[width=.8\linewidth]{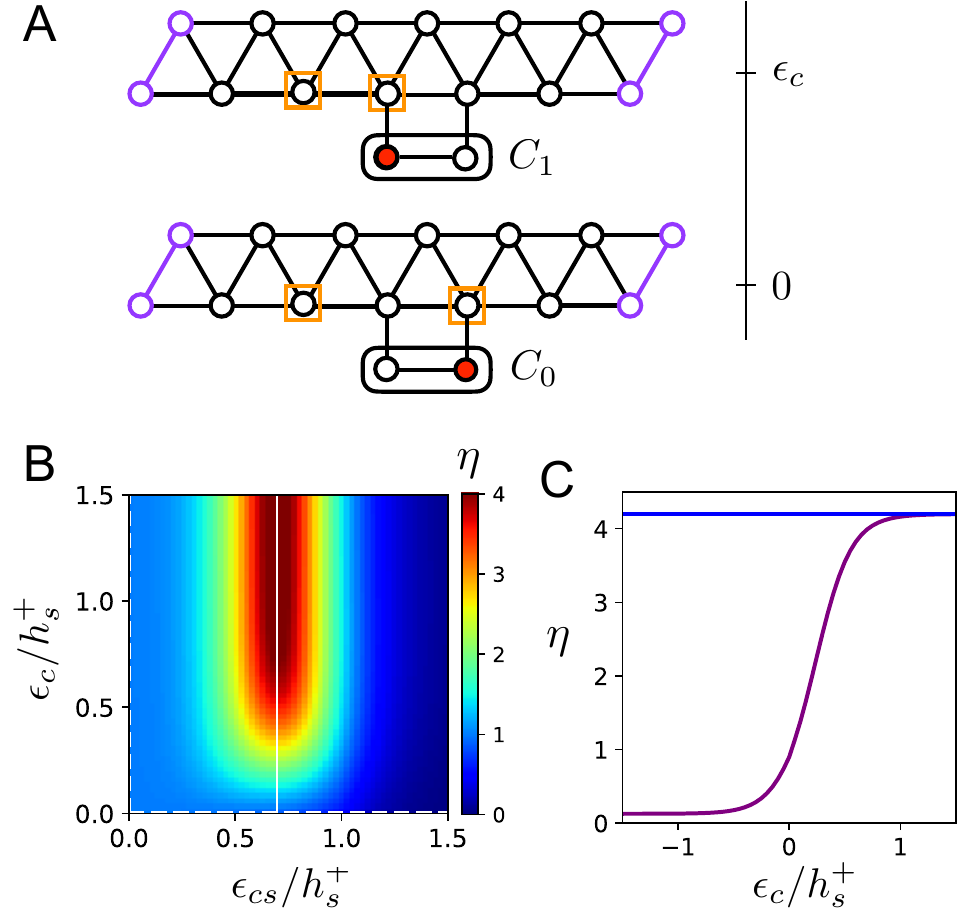}
\caption{\c Catalysis with two mobile binding sites -- {\bf A.} In this variant of the model, the position of one of two binding sites is controlled by the location of a third particle (in red) confined to two additional lattice sites. When the red particle is on the right site (state $C_0$), the distance between the binding sites is $L_c=2$. When the red particle is on the left site (state $C_1$), the distance between the binding sites is $L_c=1$, at the cost of an extra energy $\e_c$. {\bf B.} Catalytic efficiency $\eta=T_{S\to 2P}/T_{C+S\to C+2P}$ as a function of the binding energy $\e_{cs}$ and of the internal energy $\e_c$, showing an optimum when $\e_c\to\infty$. As in other figures, $h_s^+=6$ and $h_s^-=12$. {\bf C.}~Catalytic efficiency $\eta$ as a function of $\e_{c}$ for the value of $\e_{cs}$ that optimizes $\eta$ when $\e_{c}=0$ (white horizontal line in B), showing an optimum when $\e_{c}\to\infty$, i.e., when the catalyst cannot access $C_1$ and is thus effectively rigid. \label{fig:S4}}
\end{center} 
\end{figure}

We consider here an extension of the model where the distance between the two binding sites can fluctuate between $L_c=2$ and $L_c=1$. In our framework where degrees of freedom are described by particles occupying lattice sites with no two particles on the same site, this can be described by a particle confined to two extra sites (Fig.~S\ref{fig:S4}A). This is thus similar to the model of Fig.~3 except that the position of this third particle now dictates whether $L_c=2$ or $L_c=1$ without changing the binding energies, which are always $\e_{cs}$. For the sake of generality and to recover a rigid catalyst as a limit case, we also consider that the ``closed state'' with $L_c=1$ is associated with an extra energy $\e_c$. The rigid single-state catalyst then corresponds formally to the limit $\e_c\to\infty$. Numerical calculations with this system show that this limit case is indeed optimal (Fig.~S\ref{fig:S4}B-C).

This can be understood again by looking at the limit $h_s^+\gg 1$ where the dynamics is described by a Markov chain with few states,
\beq
\begin{tikzcd}[cramped]
C_0+S \arrow{r} \arrow{d}
  & C_0\.S\arrow{d}\arrow{r}\arrow{l}
  & C_0\:S^\ddagger\arrow{r}\arrow{d}\arrow{l}
  & C_0\.P\arrow{r}\arrow{l}\arrow{d}
  & C_0 \arrow{l}\arrow{d}\\
C_1+S \arrow{u}\arrow{r}
  & C_1\.S\arrow{r}\arrow{l} \arrow{u}
  & C_1\.S^\ddagger\arrow{r} \arrow{u} \arrow{l}
  & C_1\.P \arrow{u}\arrow{r}\arrow{l}
  & C_1\arrow{u}\arrow{l}\nonumber
\end{tikzcd}
\eeq
where $C_k\.P$ stands for $C_k.P+P$ and $C_k$ for $C_k+2P$.
As the step $C_1\.S\to C_1\.S^\ddagger$ involves a barrier $h_s^++\e_{cs}$ larger than the barrier for the spontaneous reaction $h_s^+$, it appears from this diagram that the extra state $C_1$ is not favorable to catalysis with this type of design.}

{\c \section*{Catalysts with a symmetric discriminative switch}

\begin{figure}[t]
\begin{center}
\includegraphics[width=\linewidth]{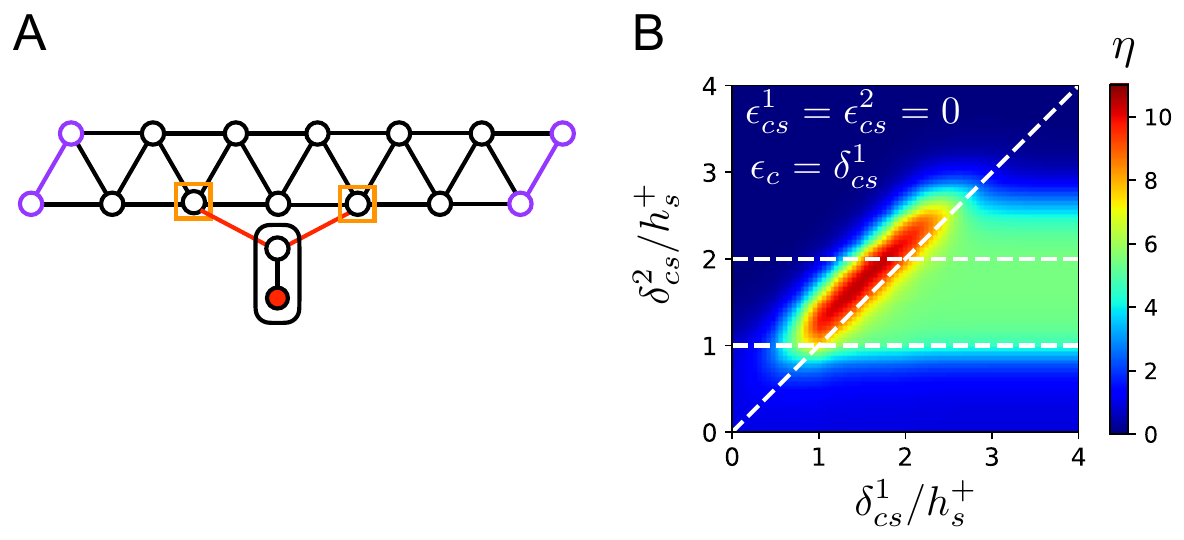}
\caption{\c Catalysis with a discriminative switch -- {\bf A.} The catalyst has an internal degree of freedom represented by the position of a third particle (in red) confined to two additional lattice sites. When the red particle is on the bottom site (state $C_0$), the interaction energy is $\e_{cs}^1$ at the first binding site and 
$\e_{cs}^2$ at the second binding site, with $\e_{cs}^1=\e_{cs}^2=\e_{cs}$ in the symmetric case. When the red particle is on the bottom site (state $C_1$), the interaction energy is $\e_{cs}^1+\d_{cs}^1$ at the first binding site and $\e_{cs}^2+\d_{cs}^2$ at the second binding site, 
with $\d_{cs}^1=\d_{cs}^2=\d_{cs}$ in the symmetric case. In addition, state $C_1$ is accessed at an energy cost $\e_c>0$. 
{\bf B.}  Catalytic efficiency for asymmetric designs as a function of $\d_{cs}^1=\e_c$ and $\d_{cs}^2$ for $\e_{cs}^1=\e_{cs}^2=0$. The horizontal dashed lines delineate the range $h_s^+<\d_{cs}^2<h_{s}^-$ ($h_s^+=6$ and $h_s^-=12$).
\label{fig:S5}}
\end{center} 
\end{figure}

\begin{figure}[t]
\begin{center}
\includegraphics[width=.9\linewidth]{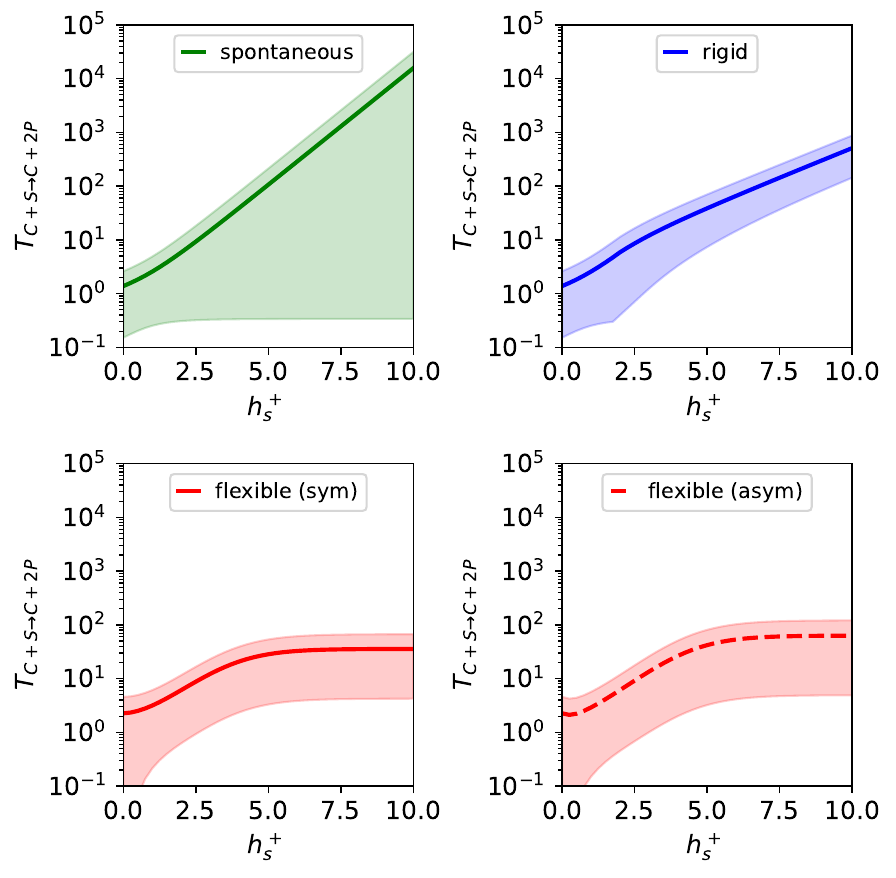}
\caption{\c Means and standard deviations of first passage times for the spontaneous reaction, an optimal rigid catalyst, a flexible symmetric catalyst and a flexible asymmetric catalyst. The parameters are exactly as in Fig.~3 where only the mean first passage times are shown. The shaded areas indicate deviations from the mean by one standard deviation. This shows that in the limit of infinite barriers $h_s^+\to\infty$ the distributions of first passage times for rigid and flexible catalysis are increasingly distinct.\label{fig:Snew}}
\end{center} 
\end{figure}

We consider here a catalyst with a discriminative switch as described in Fig.~3. We take again the limit of large energy barriers where the dynamics can be reduced to a Markov process with very few states. The path of interest is the following path marked by plain arrows,
\beq
\begin{tikzcd}
C_0+S \arrow{r}{(1)}
  & C_0\.S\arrow{d}{(2)} 
  & C_0\:S\arrow[dashed]{r}{(-1)} 
  & C_0\.P\arrow{r}{(6)} 
  & C_0 \\
C_1+S \arrow[dashed]{u}{(0)} 
  & C_1\.S\arrow{r}{(3)} 
  & C_1\:S\arrow{r}{(4)} 
  & C_1\.P \arrow{u}{(5)}
  & C_1\arrow[dashed]{u}{(0)}
\end{tikzcd}\nonumber
\eeq
where $C_{\s}+S$ indicates that $C$ is in state $\s$ and not interacting with the substrate, $C_{\s}\.S$ that the substrate occupies one binding site, $C_{\s}\:S$ that it occupies both, $C_{\s}\.P$ that one binding site is occupied by a monomer while the other is free and $C_{\s}$ that the two binding sites are free and the dimer dissociated. Each step $A\to B$ involves an activation energy $E_B-E_A$ and requiring no activation energy along the path ($E_B\leq E_A$) implies (1)~$0\leq \e_{cs}$; (2) $\e_c\leq \d_{cs}$; (3) $h_s^+\leq \e_{cs}+\d_{cs}$; (4)~$\e_{cs}+\d_{cs}\leq h_s^-$; (5) $\d_{cs}\leq\e_c$; (6) $ \e_{cs}\leq 0$. In addition, kinetic traps are avoided if the dashed arrows are also associated with negative activation energies, i.e., ($-1$) $\e_{cs}< h_s^-$, which is implied by (4), and (0) $\e_c>0$, which is an additional constraint. Assuming $0<h_s^+<h_s^-$, these different conditions are satisfied simultaneously provided
\beq
\e_{cs}=0\quad{\rm and}\quad h_s^+\leq \d_{cs}=\e_c\leq h_s^-.
\eeq
Taking for instance $\e_c=\d_{cs}=(h_s^++h_s^-)/2$, we verify that $T_{C+S\to C+2P}$ indeed does not scale exponentially with $h_s^+$ any more (Fig.~3B).}

\section*{\c Catalysts with an asymmetric discriminative switch}

{\c If relaxing the assumption that the two binding sites are equivalent, we have a total of five parameters describing the catalyst: $\e_c$, $\e_{cs}^1$, $\d_{cs}^1$, $\e_{cs}^2$, $\d_{cs}^2$.} We indicate by $C_{\s_c}\.S_k$ and $C_{\s_c}\.P_k$ that a catalyst in state $\s_c$ is bound to a single particle of the substrate at site $k$ and consider the following down-hill path:
\beq
\begin{tikzcd}
C_0+S \arrow{r}{(1)}
  & C_0\.S_1\arrow{d}{(2)} 
  & C_0\:S\arrow[dashed]{r}{(-1)} 
  & C_0\.P\arrow{r}{(6)} 
  & C_0 \\
C_1+S \arrow[dashed]{u}{(0)} 
  & C_1\.S_1\arrow{r}{(3)} 
  & C_1\:S\arrow{r}{(4)} 
  & C_1\.P_1 \arrow{u}{(5)}
  & C_1\arrow[dashed]{u}{(0)}
\end{tikzcd}\nonumber
\eeq
to which we add the following down-hill paths to prevent kinetic traps:
\beq
\begin{tikzcd}
C_1\.S_2 \arrow[dashed]{r}{(-1')}
  & C_0\.S_2\arrow[dashed]{r}{(-2')} 
  & C_0+S\\
  C_1\.P_2 \arrow[dashed]{r}{(-1'')}
  & C_0\.P_2\arrow[dashed]{r}{(-2'')} 
  & C_0\nonumber
\end{tikzcd}
\eeq
The requirements for each of the arrow to be down-hill are (1)~$0\leq \e_{cs}^1$, 
(2) $\e_c\leq \d_{cs}^1$, 
(3) $h_s^+\leq \e_{cs}^2+\d_{cs}^2$, 
(4)~$\e_{cs}^2+\d_{cs}^2\leq h_s^-$, 
(5) $\d_{cs}^1\leq\e_c$, 
(6) $ \e_{cs}^1\leq 0$, 
($-1'$) and ($-1''$) $\d_{cs}^2\leq\e_c$, 
($-2'$) and ($-2''$) $\e_{cs}^2\leq 0$,
which lead to the conditions summarized in Eq.~(7) in the main text.

\end{document}